\documentclass[numreferences]{kluwer}

\usepackage{makeidx}   
\usepackage{graphicx}  
\usepackage{multicol}  

\begin{document}
\begin{article}

\begin{opening}

\title{Type Ia Supernova models: latest developments}
\author{S. \surname{Blinnikov}\email{sergei.blinnikov@itep.ru}}
\institute{ITEP, 117218, Moscow, Russia, and MPA, Garching, Germany}
\author{E. \surname{Sorokina}\email{sorokina@sai.msu.su}}
\institute{Sternberg Astronomical Institute, 119992 Moscow, Russia, and MPA, Garching, Germany}
\runningtitle{Type Ia Supernova models}
\runningauthor{S. Blinnikov and E. Sorokina }

\begin{abstract}

Supernovae of type Ia (SNe~Ia) are very important for cosmography.
To exclude systematic effects in linking the observed light of distant SNe~Ia
to the parameters of cosmological models, one has to understand the nature
of supernova outbursts and to build accurate algorithms for predicting their
emission.\\
We review the recent
progress of modeling the propagation of nuclear flame subject to numerous
hydrodynamic instabilities inherent to the flame front. The Rayleigh--Taylor
(RT) instability is the main process governing the corrugation of the front on
the largest scales, while on the smallest scales the front propagation is
controlled by the  Landau-Darrieus instability.\\
Based on several hydrodynamic explosion models, we predict the broad-band UBVI
and bolometric light curves of SNe Ia, using our 1D-hydro code which models
multi-group time-dependent non-equilibrium radiative transfer inside SN ejecta.
We employ our new corrected treatment for line opacity in the expanding medium,
which is important especially in UV and IR bands. The results are compared with
the observed light curves. Especially interesting is
a recent  3D-deflagration
model computed at MPA, Garching, by M.~Reinecke et al.
\end{abstract}

\end{opening}

\section{Introduction}

Supernovae of type Ia (SNe~Ia) are important for cosmology due to their
brightness. They are not standard candles, but can be used for measuring
distances (i.e., for cosmography) with the help of the peak luminosity --
decline rate correlation, established by Yu.P.~Pskovskii \cite{Ps77} and
M.M.~Phillips \cite{Ph93} (see the review \cite{leib00}). The physical
understanding of the Pskovskii-Phillips relation is crucial for estimating the
validity of cosmological results obtained with SNe. To exclude systematic
effects in linking the observed light of distant SNe~Ia to the parameters of
cosmological models, one has to understand the nature of supernova outbursts
and to build accurate algorithms for predicting their emission.

This involves:
\begin{itemize}
  \item understanding the progenitors of SNe~Ia;
  \item the birth of
thermonuclear flame and it accelerated propagation leading to explosion;
  \item light curve and spectra modeling.
\end{itemize}

Although there is no doubt that an SN~Ia outburst is a result of thermonuclear
explosion, the details of the process are not yet clear.
We point out some problems which seem most important to us.

We review the recent
progress of modeling the propagation of nuclear flame subject to numerous
hydrodynamic instabilities inherent to the flame front. The Rayleigh--Taylor
(RT) instability is the main process governing the corrugation of the front on
the largest scales, while on the smallest scales the front propagation is
controlled by the  Landau-Darrieus instability.

Based on several hydrodynamic explosion models, we predict the broad-band UBVI
and bolometric light curves of SNe Ia, using our 1D-hydro code which models
multi-group time-dependent non-equilibrium radiative transfer inside SN ejecta.
We employ our new corrected treatment for line opacity in the expanding medium,
which is important especially in UV and IR bands. The results are compared with
the observed light curves. 
It seems that classical 1D thermonuclear supernova models produce 
the light curves that fit the observations not so good as the recent 
angle-averaged 3D deflagration model computed at MPA, Garching, 
by M.~Reinecke et al.~ \cite{martin}.
We believe that the main
feature of the latter model, which allows us to get the correct flux during the
first month, is strong mixing that moves the material enriched with radioactive
nickel-56 to the outermost layers of the SN ejecta.

\section{Progenitors}

There is no hope to get a thermonuclear supernova
from a normal star composed of a classical plasma.
Those stars have negative effective heat capacity and they are thermally stable.
The situation changes, if a star is made of
a degenerate matter.
The total heat capacity becomes positive,
and runaway can set in as in terrestrial explosives.
So, a progenitor of SN~Ia must be a degenerate star - a white dwarf.

A single white dwarf is unable to explode, it cools down.
But when it is in a binary system the chances to produce a supernova do appear (we need
only one in $\sim 300$ dying white dwarfs to explode in order to explain the rate
of SNe~Ia).
Even if the binary has two dead white dwarfs, it can explode because they can
merge due to emission of gravitation waves (double-degenerate, or DD scenario \cite{iben97}).
If one star in the binary is alive, the white dwarf can accrete its lost mass
and reach an instability
(single-degenerate, or SD scenario \cite{whelib73,brag90}).
It is unclear which scenario is most important, there are strong arguments \cite{kob98}
from chemical evolution that only SD is the viable one.
On the other hand, it seems that DD can produce a richer variety of SN~Ia events.
Moreover,
discoveries of intergalactic SNe~Ia \cite{bart97,galgf02} can be explained more
naturally, because a DD system may evaporate from a galaxy.
It is quite likely that both scenarios are being played, but their relative role
may change in young and old galaxies. If so, a systematic trend may appear in SNe~Ia
properties with the age of Universe, and this may have important consequences for
cosmology.

\section{Thermonuclear flames}

After merging  in DD scenario, or after the white dwarf accretes large amount of material
in SD case, the explosive instability  develops.
In principle, combustion can
propagate either in
the form of a supersonic  \emph{detonation} \cite{arn69} wave, or
as a subsonic \emph{deflagration} \cite{iich74,nomsn76} (flame).
In detonation, the unburned fuel is
ignited  by a shock front propagating
ahead of the burning zone itself.
In {deflagration}, the ignition
is governed by heat and active reactant transport, i.e.
by thermal conduction and diffusion.

\subsection{Laminar flames}

Most likely, the runaway starts as a laminar flame propagating due
to thermal conduction.

The rate of thermonuclear heating scales as
$$
\langle \sigma v_0\rangle \sim \exp-(\alpha_G/T)^{1/3}
$$
due to the Gamow's peak:
the chances to penetrate the Coulomb barrier for fast nuclei
grow, but the tail of Maxwell distribution goes down.
Here $\alpha_G$ depends strongly on nuclei charges $Z_i$:
$\alpha_G \propto Z_1^2Z_2^2 $,
thus high-$Z$ ions can fuse only at high $T$.
Small perturbations of $T$ produce huge variations
in energy production rate  since, normally, { $T \ll \alpha_G$}.

 In terrestrial flames,
the `fusion' of molecules goes with the rate:
$$
 \langle \sigma v_0\rangle \sim  \exp ( -{E_a / {\mathcal R} T}),
$$
-- the Arrhenius law of chemical burning.
Here {$E_a$} is \emph{activation energy}.
The parameter, showing the strong $T$-dependence of the heating
$$
 \mathrm{Ze}=
 {\partial \log \langle \sigma v_0\rangle   / \partial \log  T} \simeq
 { E_a / {\mathcal R} T}
$$
is called the Zeldovich number in the theory of chemical flames.
For them typically $\mathrm{Ze} \sim 10 \dots 20$.
The classical theory \cite{zfk38}
predicts the flame speed
$$
 v_\mathrm{f} \approx \mathrm{Ze}^{-1}[{v_T l_T}/ \tau_\mathrm{reac}(T_2)]^{1/2},
$$
with $T_2$ -- the temperature of burnt matter (ashes) and
$
 \tau_\mathrm{reac}(T) \propto \exp[ E_a /({\mathcal R} T)]
$.  
 In SNe, for nuclear flames,
$
 \tau_\mathrm{reac}(T) \propto \exp[\alpha_G^{1/3} /( 3 T^{1/3})]
$,
 and,
$$
 \mathrm{Ze}=
 {\partial \log \langle \sigma v_0\rangle / \partial \log  T} \simeq
  \alpha_G^{1/3} /(3 T^{1/3}),
$$
which has values very similar to terrestrial chemical flames.

A big difference with chemical flames is the ratio of heat
conduction and mass diffusion,
the Lewis number,
$
  \mathrm{Le}= (v_T l_T) / (v_D l_D)
$.
$\mathrm{Le}\sim 1$ in laboratory gaseous flames, while
$\mathrm{Le}\sim 10^7$ in thermonuclear SNe, since
heat is transported by relativistic electrons,
$v_T \sim c$, and there is almost no diffusion, $l_T \gg  l_D$.
Nevertheless, the modern computations \cite{timw92} follow
the old theory \cite{zfk38} closely.
The conductive flame propagates in a presupernova with  $v_{\rm f}$
which is too slow to produce an energetic explosion:
the ratio of $v_{\rm f}$  to sound
speed, i.e. the Mach number, Ma, is very small (see Table \ref{Tab1}).
The star has enough time to expand, to cool down, and
the burning dies completely.
So an acceleration of the flame is necessary in order to explain the
SN phenomenon. This is the main problem in current research of SNe~Ia hydrodynamics.

\begin{table}
\caption{Flame speed $v_{\mathrm f}$ and width $l_{\mathrm f}$
 in C+O \protect\cite{timw92} }
\begin{tabular}{|lllll|}
\hline\noalign{\smallskip}
 $\rho $ & $v_{\mathrm f}$ & $l_{\mathrm f}$ & $ \Delta\rho/\rho$   & Ma  \\
 $10^9$ gcc  & km/s & cm & &   \\
\noalign{\smallskip}
\hline
\noalign{\smallskip}
6  &  214 &  $ 1.8\times 10^{-5}$  &  0.10  &  $2\times 10^{-2}$ \\
1  &  36 &  $ 2.9\times 10^{-4}$  &  0.19  &  $4\times 10^{-3}$ \\
0.1  &  2.3 &  $ 2.7\times 10^{-2}$  &  0.43  &  $4\times 10^{-4}$ \\
\hline
\noalign{\smallskip}
\end{tabular}
\label{Tab1}
\end{table}

\subsection{Flame Instabilities and its Acceleration}

There is a rich variety of instabilities that can severely distort the shape
of a laminar flame.
The Rayleigh--Taylor
(RT) instability governs the corrugation of the front on
the largest scales. On the smallest scales the flame is
controlled by the  Landau-Darrieus (LD) instability.
RT, LD instabilities  and turbulence make computations difficult,
but without them a star would not explode. All these  instabilities were
considered  already by L.Landau \cite{lan44} as a means to accelerate the
flame.

Although he did not used a term `Rayleigh-Taylor instability', Landau
derived his dispersion relation with account of gravity, so RT is there!
Let $k$ be wave number, and $\Omega$ frequency of perturbations on the flame
front, the shape of perturbations being of the form $\exp(i ky+\Omega t)$.
Here we reproduce most important expressions for the
Landau dispersion relation when $g\neq 0$ and the surface tension of the 
fuel is 
$\alpha$.
The notation below follows  Landau \cite{lan44}, so
$v_1$ is the velocity of fuel,  $v_2$ the same for ashes, $j\equiv
\rho_1 v_1=\rho_2 v_2$ is the mass flux across the front.
Then the Landau dispersion relation is
$$
\Omega^2(v_1+v_2)+2\Omega k v_1v_2 +k^2v_1v_2(v_1-v_2)-kg(v_1-v_2)
 +\alpha k^3 {v_1v_2 / j}=0.
$$
Note that this relation contains RT instability: when $k$ is small and
$v_i$ is small, we have
$$
\Omega^2(v_1+v_2)-kg(v_1-v_2)  =0. 
$$
That is, if we replace $v_i$ by $1/\rho_i$ using $j=\rho_i v_i =
\mathrm{const}$, we get a `pure' RT dispersion equation.

When surface tension is not negligible, we have its stabilizing effect in RT:
$$
\Omega^2(v_1+v_2)-kg(v_1-v_2)  +\alpha k^3 {v_1v_2 / j}=0. 
$$
But this is normally important only for very short waves, and if  we have
$$
  k v_1 v_2 \gg g \; ,
$$
i.e. a case of short, but not extremely short waves,
then LD becomes important while the term with $\alpha$ can still remain
negligible.
The latter inequality means that gravity can be ignored, but it is just
$$
\frac{1}{k} \ll \ell_{\min}=\frac{v_1v_2}{g}.
$$
A fast flame runs from RT instability away, and the RT instability has no time
to develop.
But this does not mean that the flame becomes stable!
No stabilization is obtained for waves below $\ell_{\min}$, 
if the role of $\alpha$ or a thermal diffusion is negligible,
but the instability looks now not as RT, but as a `pure' LD.
Since $\epsilon=v_{1}/v_{2}<1$
 (from mass flux $j\equiv \rho_{1}v_{1}=
\rho_{2}v_{2}$ and $\rho_{\rm 1}>\rho_{\rm 2}$)
this gives a growing mode of LD instability:
$$
\Omega_\mathrm{LD}=\frac{\sqrt{\epsilon+\epsilon^2-\epsilon^3}-\epsilon}{\epsilon+1}k
v_{2}.
$$
Of course, in real life one has to use the full Landau relation, since
the subdivision into domains of RT and LD instabilities is only approximate.
In many laboratory experiments laminar flames remain stable.
Below a certain scale, so-called Markstein length, dissipative effects (like thermal
conduction) dominate and smear out all perturbations, so that the flame
may remain stable on small scales.
In the majority of experiments the flame is attached to a
burner, which sets bounds to the development of LD instability
at the Markstein scale, but some special experiments with spherical flames
do show a growth of LD instability and fractalization of the front.
Very nice examples of the action of the LD instability are
also observed in growth of bubbles in a overheated liquid.

Because of instabilities, the flame surface  becomes
wrinkled and its area
grows as
$
  S \propto \bar R^\alpha \; ,
$
with average radius $\bar R$ and $\alpha > 2$, i.e.
faster than $S \propto \bar R^2$.
In other words the surface  becomes
`fractal'. The exponent {$\alpha$}
is actually the fractal dimension, $\alpha = D_{\rm F}$.
The effective flame speed is determined \cite{woo90}
by the ratio of the maximum scale of the instability to the minimum one:
$
 v_{\rm eff}= v_{\rm f}(\lambda_{\rm max}/\lambda_{\rm min})^{D_{\rm F}-2}.
$

Proper understanding of the behavior of
flames cannot be reached if we restrict ourselves to a linear analysis
of the problem, even if we take into account dissipation.
However, good progress can be made with the help of a simple nonlinear model
for the flame.

When the vorticity is not important  it is possible to study in detail the
non-linear stage of LD instability and to find the fractal
dimension \cite{BS96}. 
Instead of $\epsilon$ we define
$$
 \eta = \Delta\rho/\rho = 1 - \rho_{\rm b} / \rho_{\rm u} = 1 -\epsilon \; .
$$
For $\rho_{1} \gtrsim 10^8 $ g/cm$^3$,
which is typical for preSN cores,
$\eta \lesssim 0.4$, since the unburned matter is very degenerate and its
thermal expansion relatively low.
For small $\eta$, it is easy to show that the flow behind the flame front
remains irrotational (to first order relative to $\eta$) if the incoming
flow is potential and to use a simple model to compute the flame propagation.
Laboratory flames in gases have  typical values of $\eta \approx 0.8$.
In this case a lot of vorticity is generated on the front
and the simple model is not applicable.

One can see that for low $\eta$ fractalization is not so pronounced
as for high density jumps.
We predict \cite{BS96},
$$
D_{\rm F}(\eta)({\rm LD}) \rightarrow 1+D_0 \eta^2,\;\mbox{ when} \;
 \eta \rightarrow 0 \;.
$$
Making a least-square fit to the power law dependence of the length
of the flame on the mean radius yields the values in the Table \ref{TabLD}.
We find that $D_0 \approx 0.3$.
\begin{table}[th]
\caption{$D_{\rm F}$ for small $\eta = \Delta\rho/\rho$.}
\begin{tabular}{lll}\hline
$\eta$  & $\eta^2$ & $D_{\rm F}$\\
\hline
0.3        & 0.09      & 1.022\\
0.35       & 0.122     & 1.039\\
0.4        & 0.16      & 1.046\\
\hline
\end{tabular}
\label{TabLD}
\end{table}

\subsection{Numerical Thermonuclear Flames}

The fractal description is good for LD instability while it remains mild, because it operates in a star on the scales from the flame thickness (a tiny fraction of
a cm) up to $\sim 1$ km. For the RT instability,
{$\lambda_{\rm max}/\lambda_{\rm min}$} is very uncertain
and the fractal dimension is uncertain too. 
(Although  a dependence of the flame fractal dimension on
the density jump across the front was found in SPH simulations of the flame
subject to RT instability \cite{BrGar95} similar to the LD case).

So a direct 3D numerical
simulation is necessary. The same is true for a low density regime of LD when
it is strongly coupled to turbulence (generated on the front, or cascading
from large RT vortices). A great progress is achieved here in several
groups \cite{HN00,martin,fritz,kho00}.
When simulating 3D turbulent
deflagrations one encounters two problems: the
representation of the thin moving surface separating hot and
cold material, and the prescription of the
local velocity $v_\mathrm{f}$ of this surface as a   
function of the large-scale flow
with a crude numerical resolution $>1$ km. One solution  is
sketched in \cite{martin}; for a different approach see \cite{kho00}.
In spite of the progress this problem cannot be treated as completely solved,
and even 1D approach may give
interesting results, especially for unusual SNe~Ia \cite{dunib01}.

\begin{figure}[ht]
  \centerline{\includegraphics[width=0.5\textwidth]{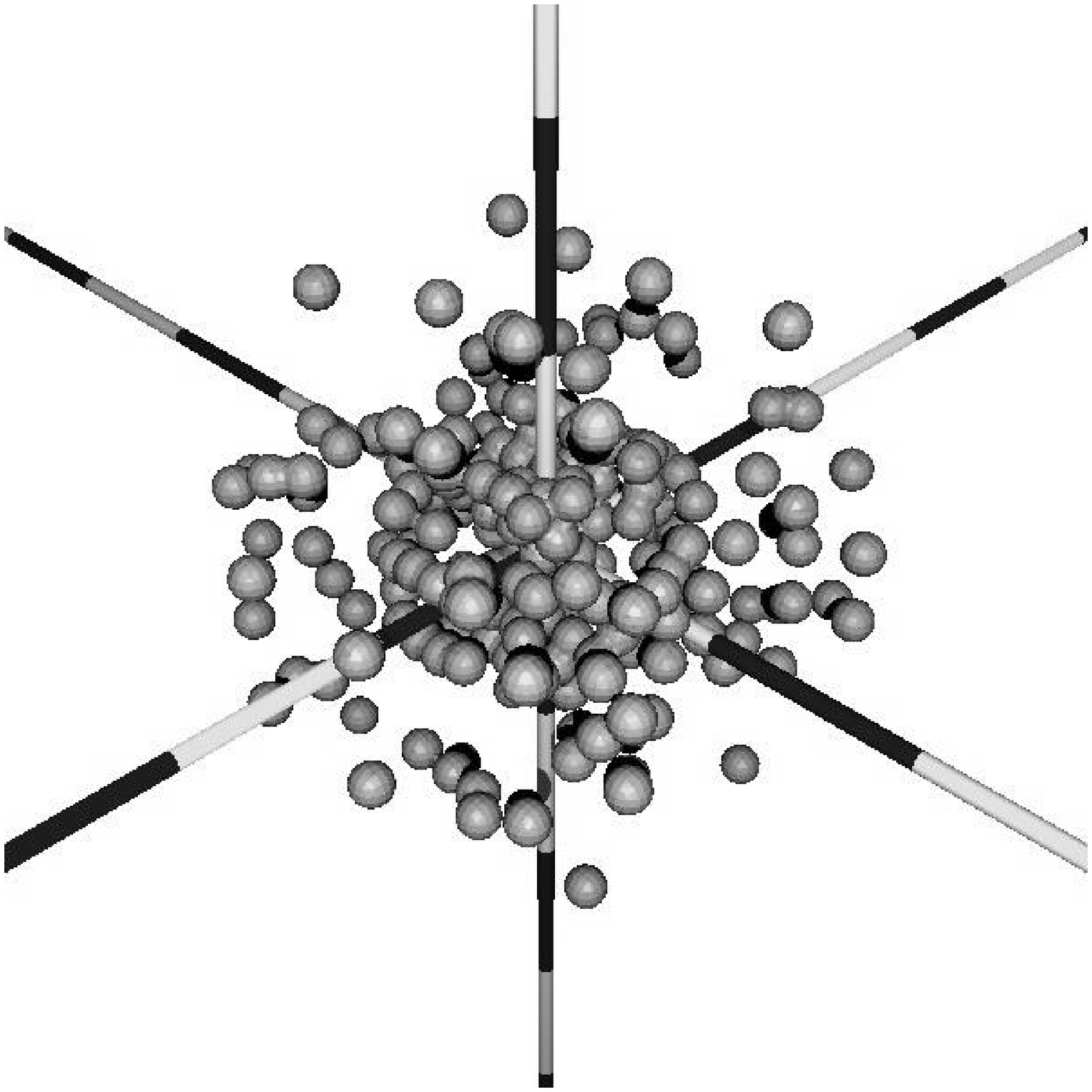} 
              \includegraphics[width=0.5\textwidth]{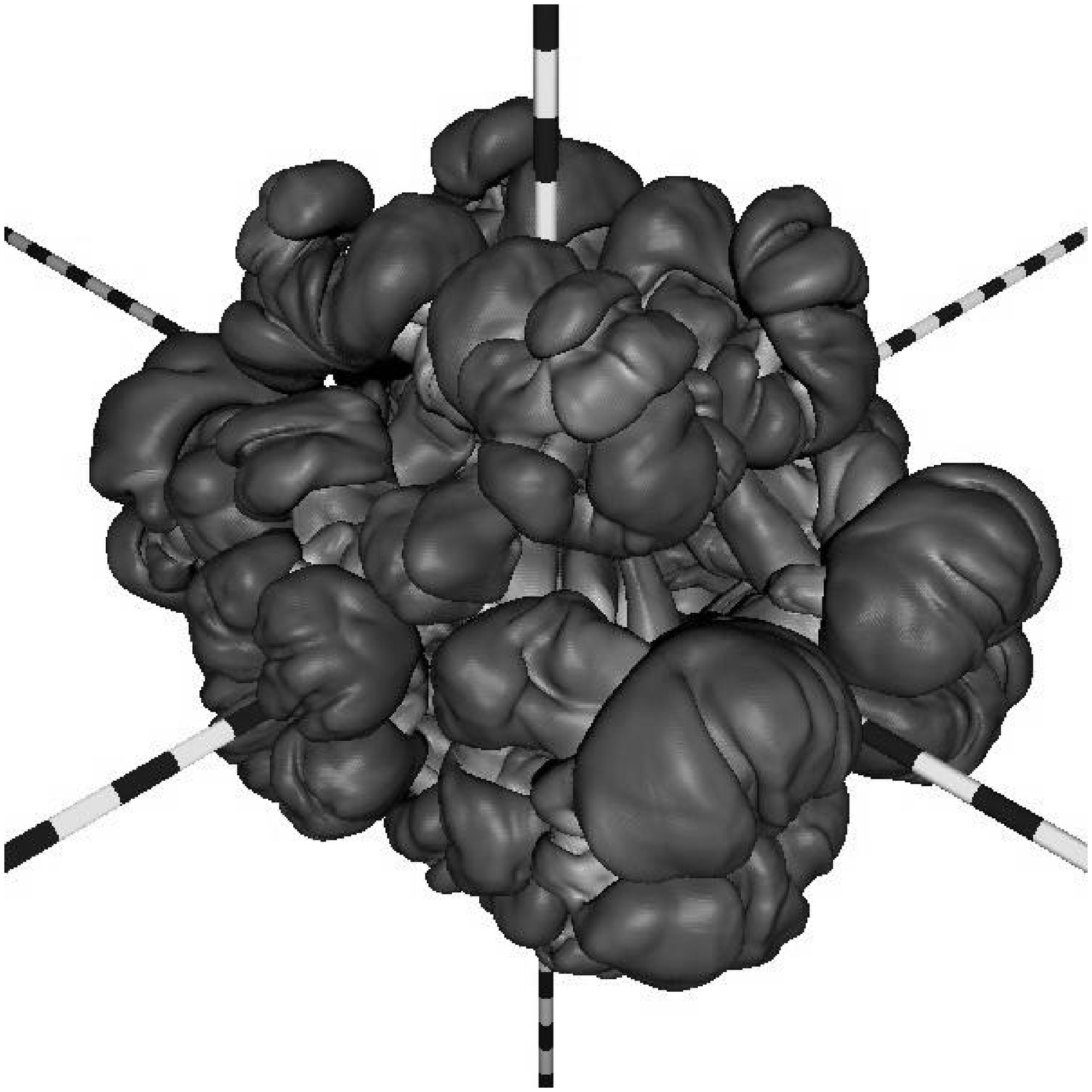}}
  \caption{Snapshots of the flame front for 3D simulations \protect\cite{MR}.
  Left panel: initial multi-point ignition.
  Right panel: flame after 0.5 seconds}
  \label{MRbigsim}
\end{figure}

\section{Light curves of SNe~Ia}

\subsection{Basics}

At the moment,
there are many models of thermonuclear
explosion of a star, that lead to the event we know as SN~Ia.
Only a few parameters,
such as kinetic energy and total $^{56}$Ni
production, can be derived directly from the modelling of the explosion
and compared with the observational values.
The subsequent evolution of the exploded star gives us 
much more possibilities to compare models and to decide which one fits
observations better by reproducing more details in SN~Ia light curves
and spectra.
We will focus here on the broad-band UBVI and bolometric light curve
computations for SN~Ia models.

There are several effects in SNe physics which lead to difficulties
in the light curve modelling of any type of SNe.
For instance, an account should be taken correctly for deposition
of gamma photons produced in decays of radioactive isotopes, mostly $^{56}$Ni
and  $^{56}$Co.
After being emitted, gamma photons travel through the ejecta and can
finish up in either thermalization or in non-coherent scattering processes.
To find this one has to solve
the transfer equation for gamma photons together with hydrodynamical
equations.
Full system of equations should involve also radiative transfer equations
in the range from soft X-rays to infrared for the expanding medium
\cite{blinn98}.

There are millions of spectral lines that form SN spectra, and it is
not a trivial problem to find a convenient way how to treat them
even in the static case.
The expansion makes the problem much more difficult to solve:
hundreds or even thousands of lines give their input into emission and
absorption at each frequency.

On the first glance,
the modelling of SNe~Ia seems easier than that
of other types of supernovae, since the hydro part is very simple:
coasting stage starts very early, there are no shocks and no
additional heating from them.

\begin{figure}[ht]
  \centerline{\includegraphics[width=0.6\textwidth]{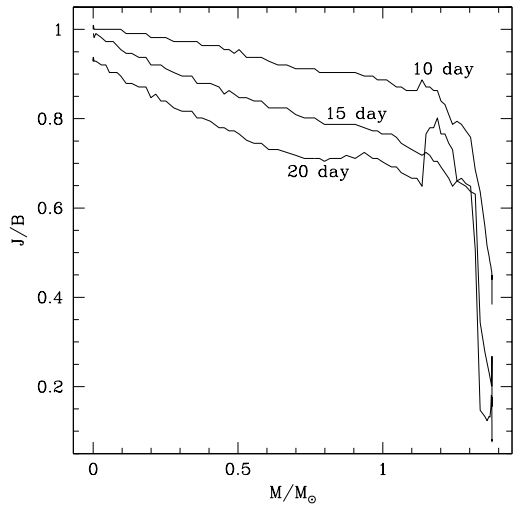}}
  \caption{The distribution of the ratio of actual radiation energy density
to the blackbody one at the local gas temperature inside the ejecta
at different times.
It demonstrates how radiation is decoupled from
matter.}
 \label{figJBM}
\end{figure}

On the other hand, much more difficulties arise in the radiation part.
SNe~Ia becomes almost transparent in continuum at the age of a few weeks.
This means that NLTE effects are stronger than for other types of supernovae.
Radiation is decoupled from
matter within the entire SN~Ia ejecta even before maximum light
(which occurs around the 20th day after explosion), see  Fig.~\ref{figJBM}
and also \cite{EPThn}.
In this case one cannot ascribe the gas temperature, or any other
temperature, to radiation,
since SN~Ia spectrum differs strongly from a blackbody one.
Instead, one has to solve a system of time-dependent transfer equations
in many energy groups,
with an accurate prescription for treatment of a huge number of spectral lines,
which are the main source of opacity in this type of SNe \cite{BHM,PE00}.

Recently, powerful codes appear aimed to attack
a full 3D time-dependent problem of SN~Ia light \cite{hoeflich}.
Yet there are some basic questions, like averaging the line opacity
in expanding media, that remain controversial.

We present theoretical UBVI- and bolometric light curves of SNe~Ia 
for several explosion models, computed with our multi-group 
radiation hydro code.
We employ our new corrected treatment for line opacity in the expanding
medium.
The results are compared with observed light curves.

\subsection{Method}

We compute broad-band UBVI and bolometric light curves of SNe~Ia with a
multi-energy radiation hydro code {\sc STELLA}.
Time-dependent equations for the angular moments of intensity  in fixed
frequency bins are coupled to Lagrangean hydro equations and solved
implicitly \cite{blinn98}.
Thus, we have no need to ascribe any temperature to the radiation:
the photon energy distribution may be quite arbitrary.

\begin{figure}[ht]
  \centerline{\includegraphics[width=0.8\textwidth]{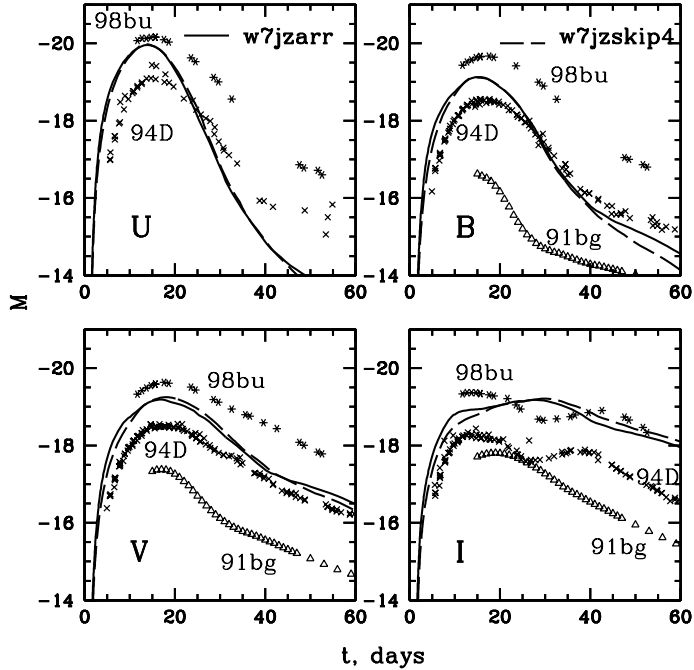}}
  \caption{UBVI light curves of the model W7 with two opacity approximations:
Eastman--Pinto {\it (dashed)} and our new approach {\it (solid)}.
UBV bands remain almost unaffected, while I--band shows now
a structure more  similar to what one can observe for real SNe~Ia,
though not identical.
Crosses, stars and triangles show the light curves
for three observed SNe~Ia.}
  \label{figLC}
\end{figure}

While radiation is nonequilibrium in our approximation, ionization and
atomic level populations are assumed to be in LTE.
The effect of line opacity is treated as an expansion opacity according to
Eastman \& Pinto \cite{EP} and to our new
recipes \cite{SBring02}.
We have compared the results and found that infrared bandpass (in which the
ejecta are most transparent) is more sensitive to the treatment of
opacity than UBV, so one should be very careful
on this point. See Fig.~\ref{figLC} for the comparison of SN~Ia light curves
calculated with these two approaches.
To simulate NLTE effects we used the approximation of the
absorptive opacity in spectral lines.
NLTE results~\cite{BHM} and ETLA approach~\cite{PE00} demonstrate that fully 
absorptive lines gives us an acceptable approximation of NLTE effects.

We treat gamma-ray opacity as a pure absorptive one, and solve the
$\gamma$-ray transfer equation in a one-group approximation 
following~\cite{SSH}.
The heating by decays  $^{56}$Ni  $\to$ $^{56}$Co$\to$ $^{56}$Fe is
taken into account.

In the calculations of SN~Ia light curves we use up to 200 frequency bins and up
to $\sim 400$ zones as a Lagrangean coordinate.

\subsection{Light curves for presupernova models}

In our previous analysis we have studied two Chandrasekhar-mass models:
the classical deflagration
model W7  \cite{W7} and the delayed detonation one DD4 \cite{DD4}, as well as
two sub-Chandrasekhar-mass models: helium detonation model LA4 \cite{livne}
and low-mass detonation model  WD065 with low $^{56}$Ni production 
\cite{pilar}, which was constructed for modelling subluminous SNe~Ia events,
such as SN~1991bg. All those models were simplified spherically-symmetrical
(1D) ones.

\begin{figure}[ht]
  \centerline{\includegraphics[width=0.8\textwidth]{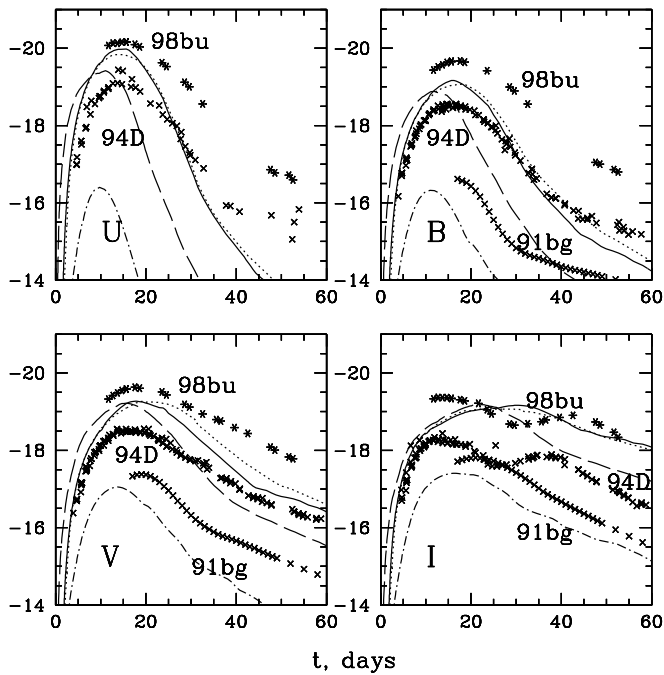}}
  \caption{UBVI light curves for 4 1D models (W7 -- solid line, DD4 -- dots,
LA4 -- dashes, WD065 -- dash-dots).
  Crosses, stars and triangles show the light curves
for three observed SNe~Ia.}
  \label{fig1DLCs}
\end{figure}

The UBVI light curves of 1D models are shown in Fig.~\ref{fig1DLCs}.
The Chandrasekhar-mass models demonstrate almost identical light curves.
The sub-Chandrasekhar-mass ones are more different.
WD065 has almost similar element distribution as Chandrasekhar-mass
models, and the shape of its light curve is in principle the same as that of
W7 and DD4.
It is just much dimmer due to an order of magnitude less $^{56}$Ni
abundance.

LA4 is very different from any other model, since the explosion there
started on the surface of a white dwarf, not in the center,
as for every other model, so there is a $^{56}$Ni layer near the surface
in LA4.
This feature explains why the model is essentially bluer
than all three others.
During first weeks UV quanta are locked inside the models
with $^{56}$Ni in the center and come out only if they are splitted
into several redder photons, while in LA4 model they can travel outside
from the surface freely.
Unfortunately, LA4 seems much bluer than real SNe~Ia, its rise
and decline rates are also too fast, so centrally ignited models look better
from the observational point of view.

\begin{table}
\caption{Parameters of SN Ia models}
\label{models}
\begin{tabular}{llllll}
\hline
Model & DD4 & W7 & LA4 & WD065 & MR \\
\hline
$M_{\rm WD}{}^{\rm a}$      & 1.3861 & 1.3775 & 0.8678 & 0.6500  & 1.4 \\
$M_{{}^{56}{\rm Ni}}{}^{\rm a}$ & 0.63 & 0.60 & 0.47   & 0.05  & 0.42 \\
$E_{51}{}^{\rm b}$ & 1.23   & 1.20   & 1.15   & 0.56  & 0.46\\
\hline
\multicolumn{5}{l}{${}^{\rm a}$in $M_\odot$} \\
\multicolumn{5}{l}{${}^{\rm b}$in $10^{51}$~ergs~s${}^{-1}$}
\end{tabular}
\end{table}

Recently, an interesting and much more sophisticated 3D deflagration model by
M.Reinecke et al. \cite{martin} (MR, hereafter) has appeared. Working in more
dimensions the authors have to involve less free parameters than was needed
in 1D simulations, so they get their  model
almost from the ``first principles''.
The main features of the 3D model are compared with the ones of classical
1D models in the Table \ref{models}.

\begin{figure}[ht]
  \centerline{\includegraphics[width=0.8\textwidth]{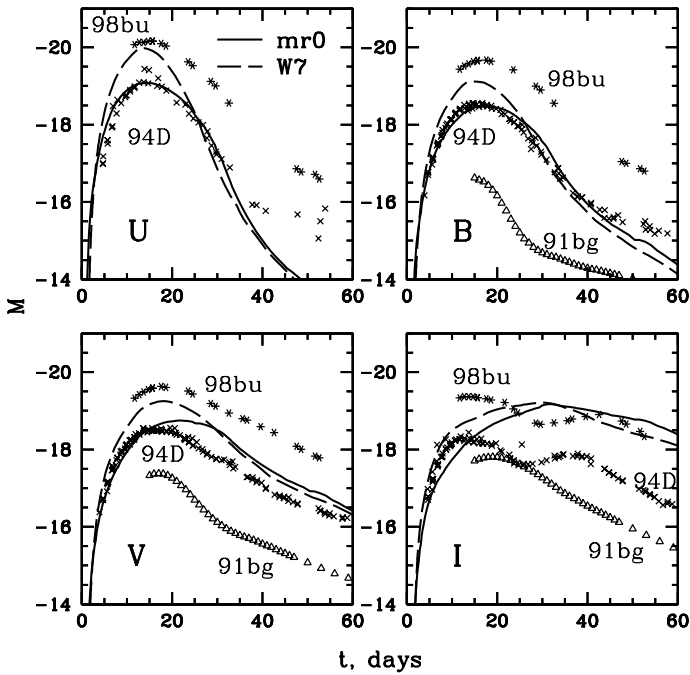} 
}
  \caption{UBVI light curves for the 3D (MR; solid) and 1D (W7; dashed) models.
  Crosses, stars and triangles show the light curves
for three observed SNe~Ia.}
  \label{figw7mr0}
\end{figure}

\begin{figure}[ht]
  \centerline{\includegraphics[width=0.5\textwidth]{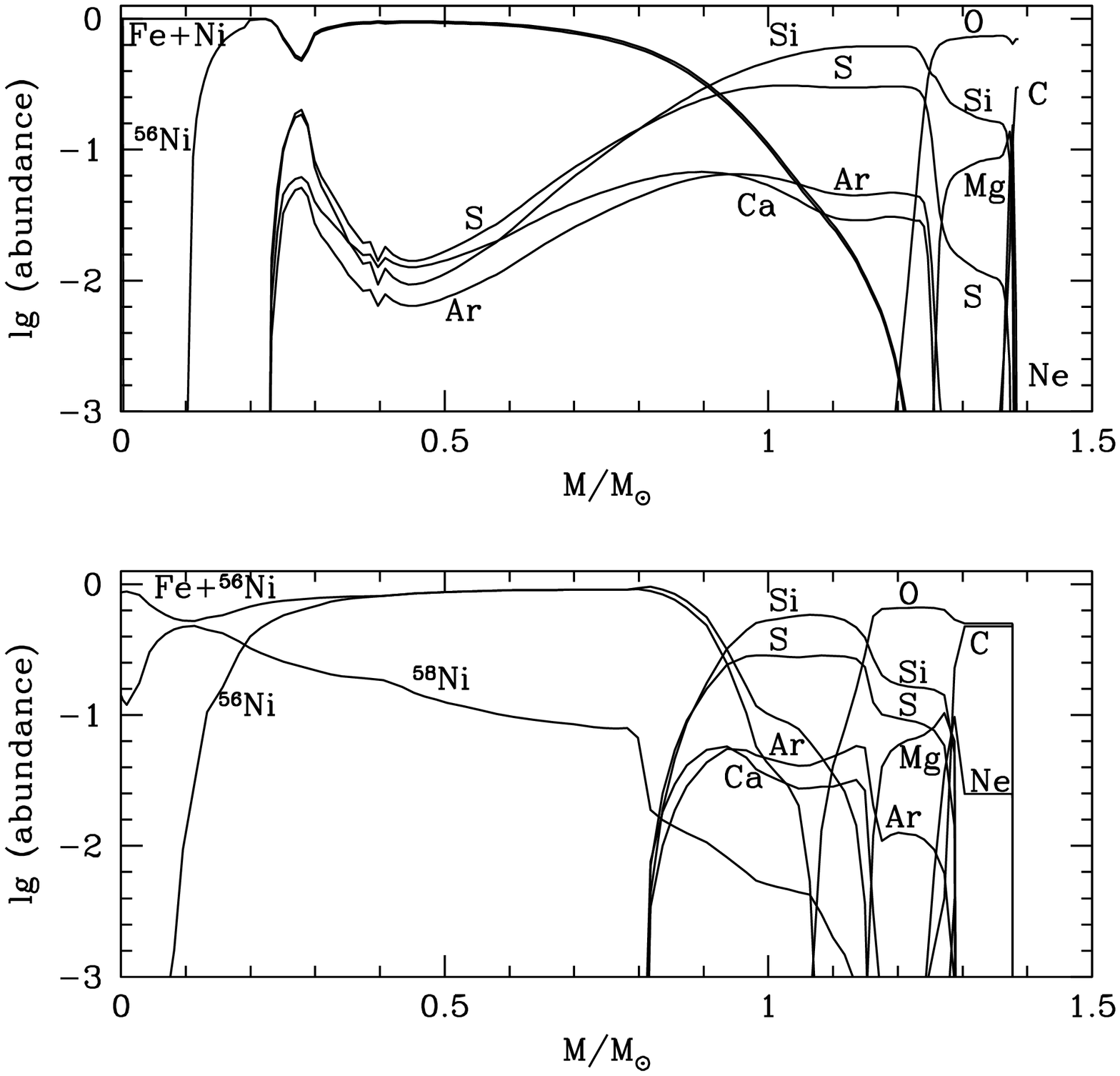} 
              \includegraphics[width=0.5\textwidth]{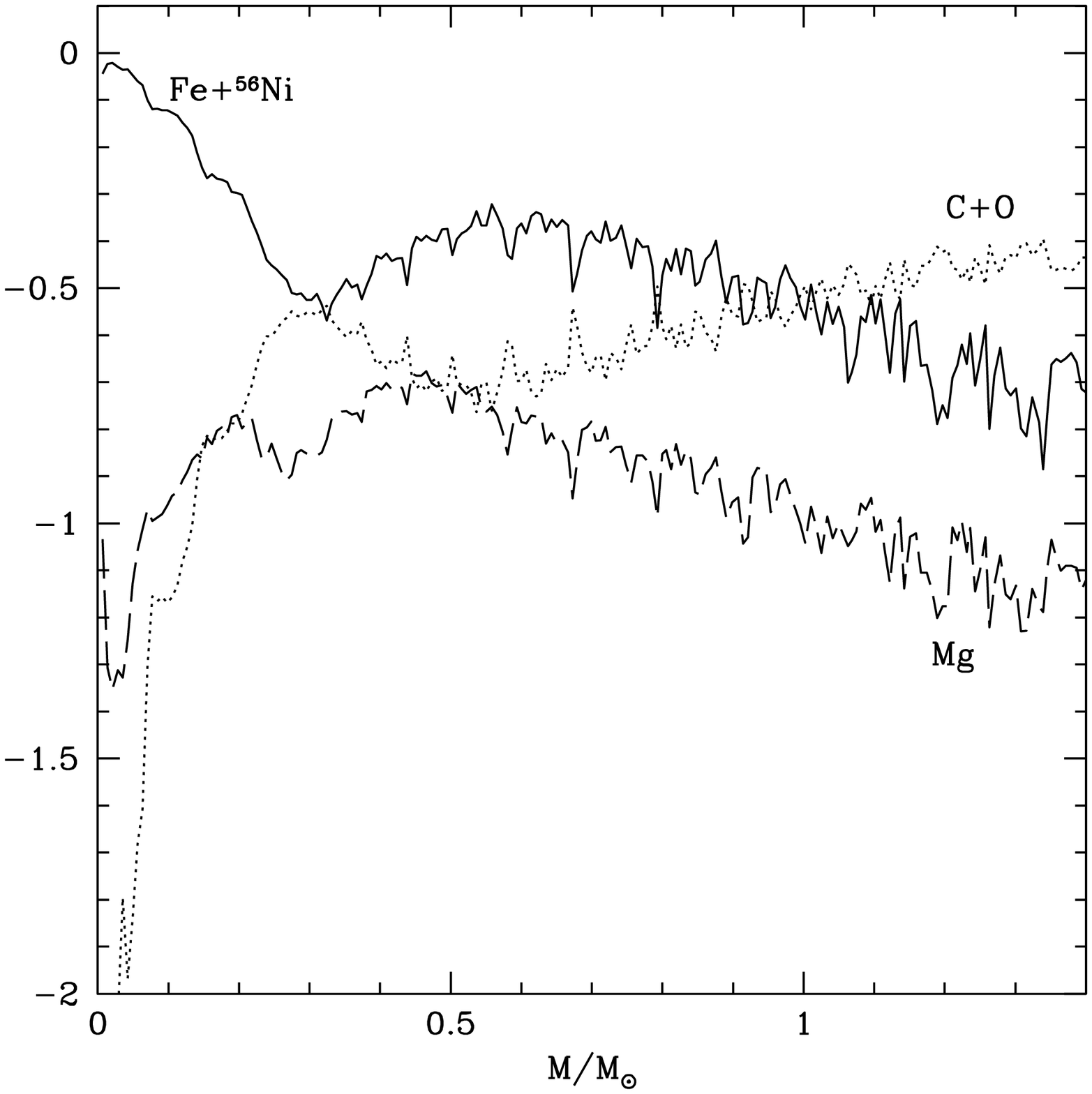}} 
  \caption{Element distributions for DD4 (top left) W7 (bottom left) and MR
  (right) models.}
  \label{figchem}
\end{figure}

From the first glance it seems that the light curves for the models which
differ so much could not be similar.
Nevertheless, Fig.~\ref{figw7mr0}
demonstrates that they are similar in many details.
The possible reasons for this can be understood if one has a look
at the element distribution over the ejecta.
The compositions for W7 and MR models are shown in Fig.~\ref{figchem}.

At the moment of our light curve computation the full nuclide  yields for MR
were not yet obtained.
Therefore, the model consisted of the elements, which were chosen
as representative examples for the energy release calculation, namely, ``Fe''
for iron group elements that were divided onto 80\% of $^{56}$Ni and 20\%
of $^{56}$Fe, ``Mg'' for intermediate mass elements, and unburned C and O
in equal proportion.

The instabilities that have developed in the 3D model were not supposed
to be so huge in approximate 1D models of explosion.
This has led to the differences in the nickel distribution over the ejecta:
it is mixed to the outermost layers in the 3D model.
These layers become much more opaque than in the 1D models, and,
despite having less than a half of kinetic energy, the 3D model has a
photospheric velocity comparable to that in the 1D models.
It is probably still a bit too low, so the light curve is wider,
since the ejecta
expand a bit slower, and photons are locked inside them for a bit longer time.
The broad-band light curves for MR model fit the observations
of one of the typical SN~Ia, SN~1994D, in $U$ and $B$ bands surprisingly well,
while classical
1D models, such as  W7 and  DD4, show faster decline in the optics
than it is observed.

\begin{figure}[ht]
  \centerline{\includegraphics[width=0.5\textwidth]{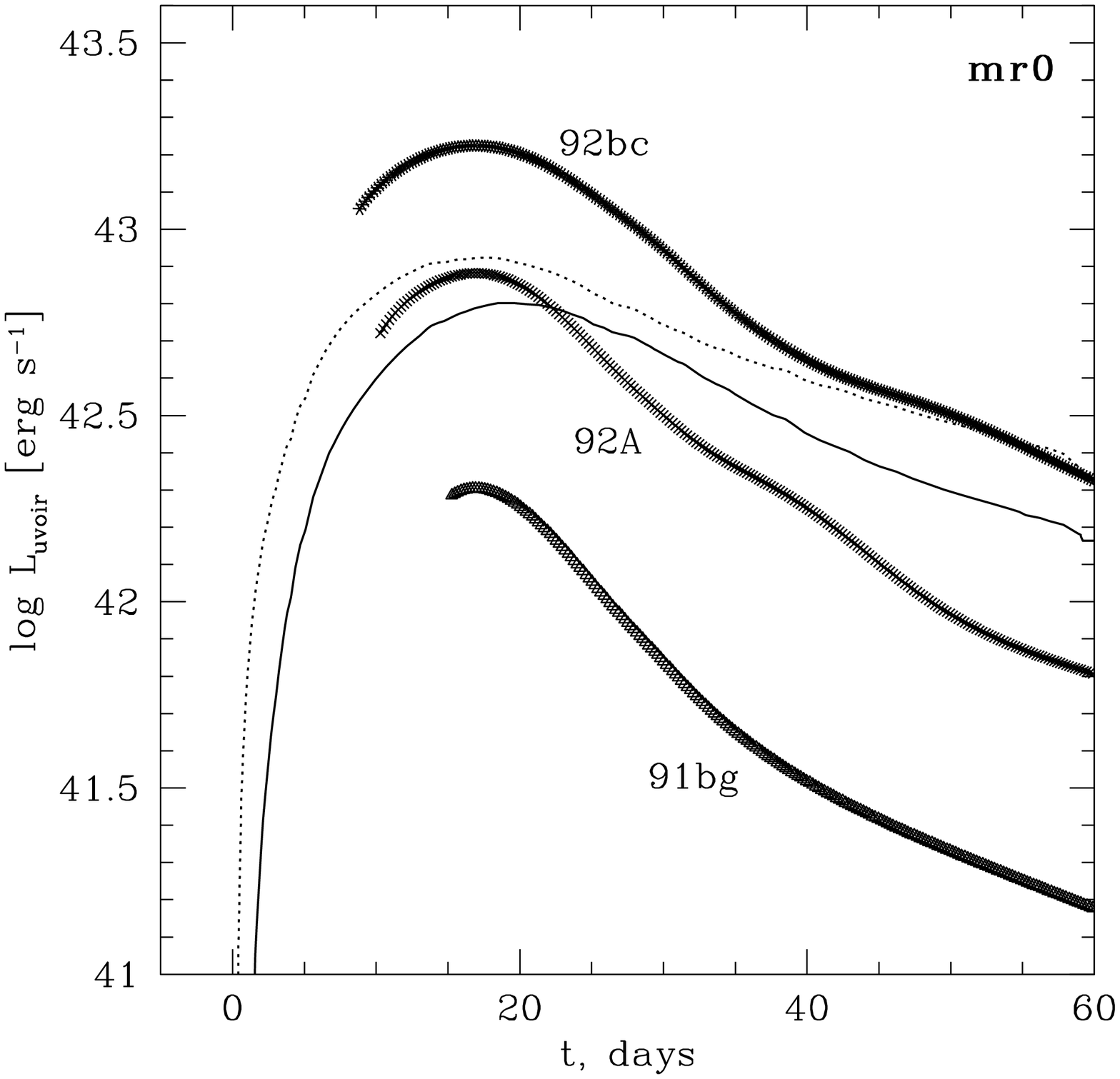} 
              \includegraphics[width=0.5\textwidth]{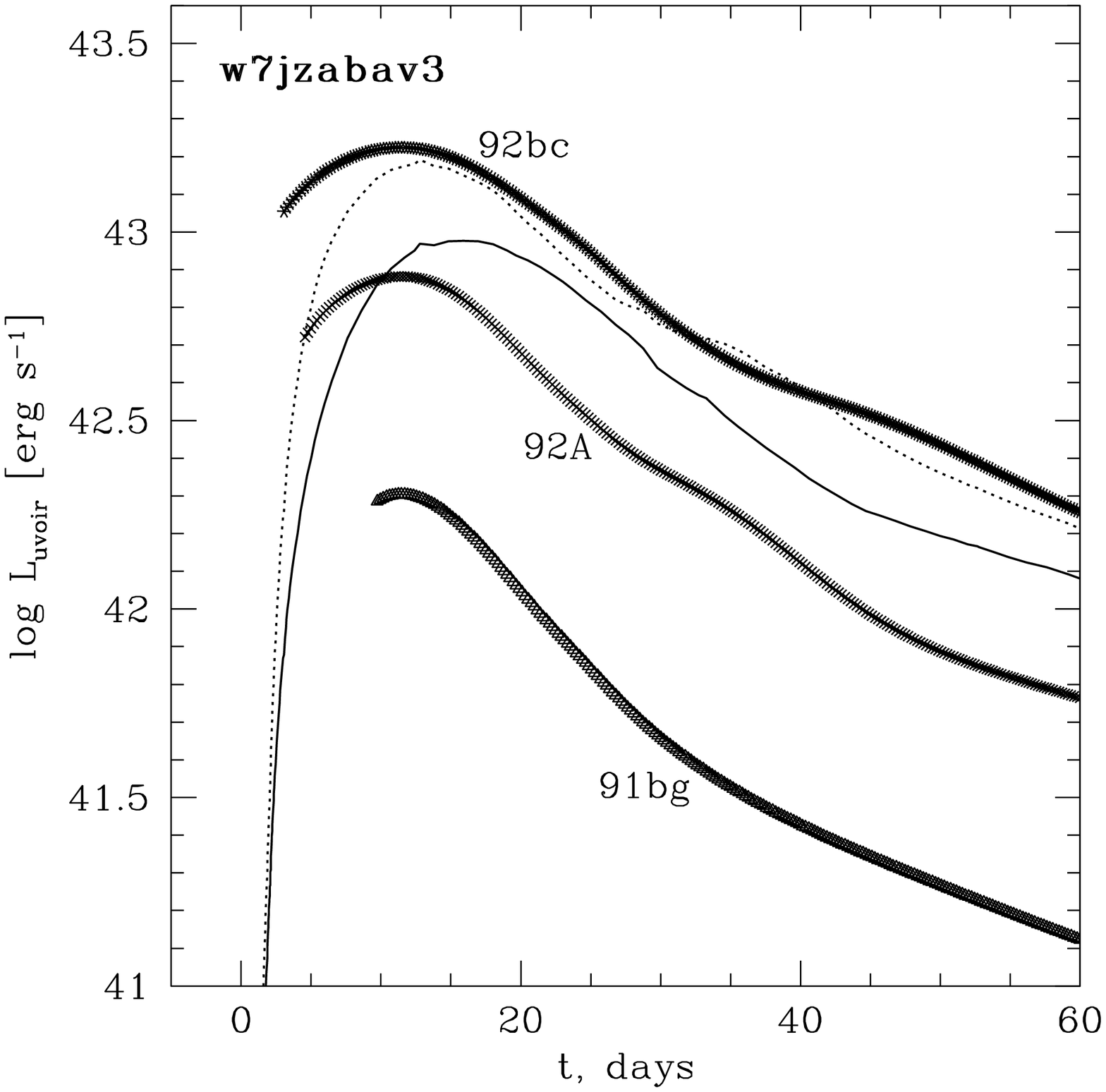}}
  \caption{Total (thin dotted) and {\it UVOIR} (thick solid)
bolometric luminosities
of MR (left panel) and W7 (right panel) models compared with observations
\protect\cite{CLV}.}
  \label{figbol}
\end{figure}

Unfortunately, the bolometric light curve for MR model
(Fig.~\ref{figbol})
is somewhat too slow.
The ejecta must expand with higher speed to let photons to diffuse out
faster.
This means that the total energy must be larger to fit
the bolometric observations.
The bolometric light curve of W7 is also shown in the Fig.~\ref{figbol}.
It demonstrate much more acceptable decline rate,
though the behaviour of the light curve before maximum light seems better
in MR model.

The model MR is not final.
The work on getting a new, more energetic 3D model is in progress at the MPA
supernova group.
It seems that such a model should still be as much mixed as MR.
Then one could expect that it would fit the observations of SN~Ia bolometric
light curves as well as the broad band ones.

There is also another reason which allows us to believe that the MR model
is better than the classical 1D models.
We calculated in details the X-ray emission of Tycho SNR, which is believed
to be the remnant of SN~Ia.
The code we use takes into account the time-dependent
ionization and recombination.
We have compared the computed X-ray spectra and
images in narrow filter bands with XMM-Newton observations of the Tycho.
Our preliminary results \cite{tycho} show
that all Chandrasekhar-mass models produce
similar X-ray spectra at the age of Tycho, but they differ strongly
in the predictions of how the remnant should look like in the lines
of different ions due to very different distribution of elements in the ejecta
for 1D and 3D models.
We found that W7 and DD4 models
produce rather wide ring in Fe lines, while it is narrow for MR model.
The image for the latter model is very similar to what is observed.

We believe that the main feature
of this model which allows us to get correct radiation during the first month,
as well as after a few hundred years, when an SNR forms, is 
strong mixing
that pushes material enriched in iron and nickel to the outermost layers
of SN ejecta.

\section{Conclusions}

There are many points which require attention in research of SNe~Ia:
progenitors;
burning regimes (that may change with the age of Universe \cite{ourLC}).
The physical understanding of the Pskovskii-Phillips is not yet achieved
(probably it will be reached when the burning will be modeled completely
from the first principles, because too many parameters enter in light curve
computations).

The new 3D SN~Ia model MR \cite{martin} is very appealing.
Yet it is not a final one: a detailed post-processing of nucleosynthesis 
changes the composition.
It has been  done
very recently  \cite{claudia}, and it is not yet
checked in the light curve calculation.
Our light curve computations are also preliminary, since
more work is needed on the expansion opacity.
Hopefully, none of the required improvements will spoil the light curve
of this model and its X-ray spectra on the SNR stage,
since the specific qualities of the model can be primarily explained
by the enrichment of the outermost layers of SN ejecta by  Fe and Ni.

The SN light curve modelling still has a  lot
of physics to be added, such as a 3D time-dependent
radiative transfer, including as much as possible of NLTE effects, which are
especially essential for SNe~Ia  \cite{hoeflich}.
All this will improve our understanding of thermonuclear supernovae.

\subsection*{Acknowledgements}

The authors are grateful to Wolfgang Hillebrandt and
to Stan Woosley for their hospitality at MPA and UCSC, respectively,
and to Bruno Leibundgut for providing us with the data \cite{CLV} in
electronic form.
The work is supported in Russia by RFBR grant 02-02-16500,
in the US, by NASA grant NAG5-8128 and in Germany, by MPA visitor program.

\end{article}

\end{document}